\documentclass[twocolumn,twoside,slac_two]{revtex4}
\pdfoutput=1
\usepackage[pdftex]{graphicx}
\usepackage{epstopdf}
\usepackage{fancyhdr}
\begin{document}

\title{Balloon-borne gamma-ray telescope with nuclear emulsion}

%

\author{Satoru Takahashi for the Emulsion Gamma-ray Telescope Group}
\affiliation{Nagoya University, Nagoya 464-8602, Japan}

%
%

\begin{abstract}
 By detecting the beginning of electron pairs with nuclear emulsion, precise gamma-ray direction and gamma-ray polarization can be detected. With recent advancement in emulsion scanning system, emulsion analyzing capability is becoming powerful. Now we are developing the balloon-borne gamma-ray telescope with nuclear emulsion. Overview and status of our telescope is described.

\end{abstract}

\maketitle

\thispagestyle{fancy}


\section{Introduction}
 The observation of high energy cosmic gamma-rays provides us direct knowledge of high energy phenomena in the universe. Fermi Gamma-ray Space Telescope with Large Area Telescope (Fermi LAT) was launched on 2008 \cite{FermiLAT}. Large scale observation has been achieved since CGRO/EGRET (launched on 1991) \cite{EGRET}. Fruitful results are being obtained in the observation of high energy cosmic gamma-rays. Further precise observations will provide us much new knowledge.

 The interaction of high energy gamma-rays with matter is dominated by electron pair creation process. Electron pair has the information of gamma-ray direction, energy, arrival timing and polarization. By suppressing multiple coulomb scattering and detecting the trajectory of electron pairs precisely, precise gamma-ray direction and gamma-ray polarization are detected.

 Nuclear emulsion is a powerful tracking device that can record 3 dimensional charged particle tracks with precise position resolution ($<1\mu$m). Several prominent observations have been performed with nuclear emulsion, i.e. discovery of $\pi$ meson \cite{Powell}, discovery of charmed meson \cite{Niu} and the first observation of tau-neutrino interactions \cite{NuTau}. For experiments, we use emulsion film (or plate) that has emulsion layers coated on both sides of plastic base. By detecting beginning of electron pairs with emulsion film, precise gamma-ray direction and gamma-ray polarization can be detected. We are developing the gamma-ray telescope consisting of emulsion films (emulsion gamma-ray telescope).

\section{Emulsion gamma-ray telescope}
\begin{figure}[htbp]
\begin{center}
\includegraphics[scale=0.3]{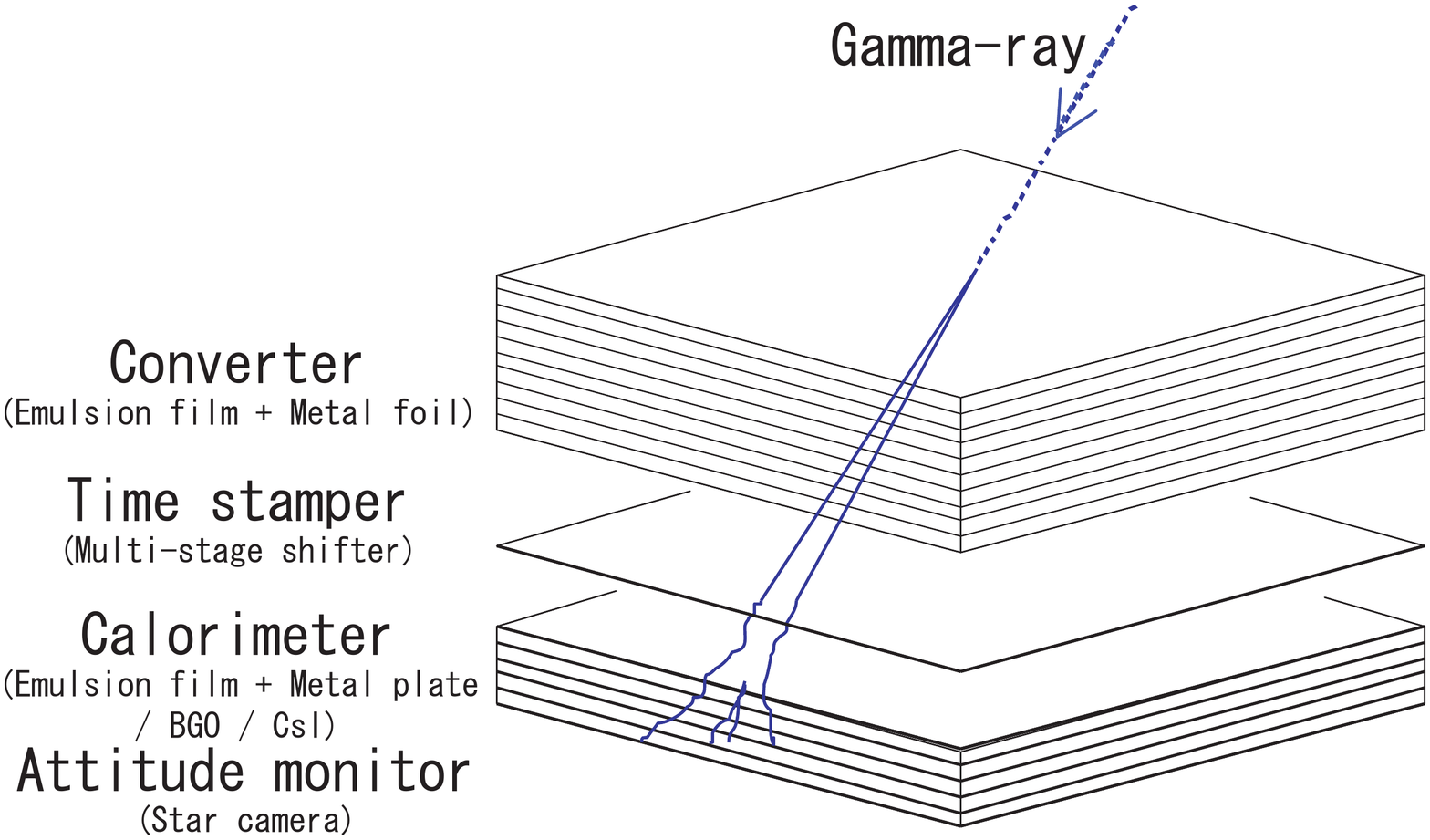}
\caption{The schematic view of emulsion gamma-ray telescope}
\label{EmulsionGammaRayTelescope}
\end{center}
\end{figure}

 Figure \ref{EmulsionGammaRayTelescope} shows the schematic view of emulsion gamma-ray telescope. Emulsion gamma-ray telescope consists of the converter, the time stamper, the calorimeter and the attitude monitor. By balloon flight with the telescope, the observation is performed. The converter consists of the stack of emulsion films and metal foils. Beginning of electron pair is detected at the converter. The time stamper consists of multi-stage shifter. Multi-stage shifter is new time stamp method for the emulsion, which is described below. The time stamper gives the time stamps to the converter events. The calorimeter consists of the stack of emulsion films and metal plates (or BGO or CsI). Gamma-ray energy above GeV is measured at the calorimeter by measuring electro-magnetic shower. Gamma-ray energy below GeV is measured at the converter by measuring multiple coulomb scattering. The Attitude monitor consists of the star camera. By combining the attitude monitor information and the event timing, gamma-ray direction to the celestial sphere is determined.

\section{Performance}
 Table \ref{Performance} shows the performance of emulsion gamma-ray telescope. Emulsion gamma-ray telescope has high angular resolution, is expected to have polarization sensitivity and has no dead time. Photon statistics from an object depend on the aperture area $\times$ the observation time. Area $\times$ time is limited by emulsion analyzing capability and balloon flight chance. For the emulsion analyzing capability, we are developing high speed automatic emulsion scanning system. With current scanning systems, 2.6 m$^2$$\cdot$flight with 100 emulsion films is possible in a year. With future scanning systems, 81 m$^2$$\cdot$flight with 100 emulsion films is possible in a year. Since the flight duration is 150 hours (6.25~days), 16 m$^2$$\cdot$day with current scanning systems and 507 m$^2$$\cdot$day with future scanning systems can be obtained, respectively. Our observation requires long duration flights and repeating flights.

\begin{table}[htbp]
\begin{center}
\caption{The performance of emulsion gamma-ray telescope}
\begin{tabular}{|l|c|}
\hline Angular resolution @ 100MeV        & 0.57 deg                                \\
\hspace{2.65cm}           @ 1GeV          & 0.08 deg                               \\
\hline Energy range                       & 10 MeV - 100 GeV                    \\
\hline Polarization sensitivity           & Expected                 \\
\hline Aperture area                      & $>$ 1 m$^2$                             \\
\hline Field of view                      & $>$ 1.6 sr   \\
\hline Dead time                          & No dead time                        \\
\hline Area $\times$ Time $^a$ (current$^b$)       & 16 m$^2$$\cdot$day                        \\
\hspace{2.1cm}           (future$^c$)       & 507 m$^2$$\cdot$day                          \\
\hline
\end{tabular}
\\Notes : $^a$ As the flight duration 150 hours, $^b$ Current scanning systems, $^c$ Future scanning systems
\label{Performance}
\end{center}
\end{table}

\subsection{Micro Segment Chamber}
 We started balloon experiment with recent emulsion techniques at Sanriku on 2004 \cite{MSC}. The shifter which has the mechanism shifting emulsion films was introduced to Micro Segment Chamber to give the time resolution to tracks recorded in emulsion film. The shifter performed to distinguish tracks recorded at each altitude. Electro-magnetic component down to 10~GeV at level flight was systematically detected by detecting electro-magnetic shower.

\subsection{Automatic emulsion scanning system}
 We are developing automatic emulsion scanning system \cite{TS}. Currently, 5 new scanning systems are running constantly \cite{SUTS}.  With current scanning systems, the total scanning speed of 600~cm$^2$/hour can be achieved. Aperture area 2.6~m$^2$ with 100 emulsion films can be analyzed in a year. The development is being started for future scanning system with higher speed. Total scanning speed with 2 systems was designed for 18400~cm$^2$/hour. Aperture area 81~m$^2$ with 100 emulsion films can be analyzed in a year. Thus emulsion analyzing capability is becoming powerful.

\subsection{Angular resolution}
\begin{figure}[htbp]
\begin{center}
\includegraphics[scale=0.3]{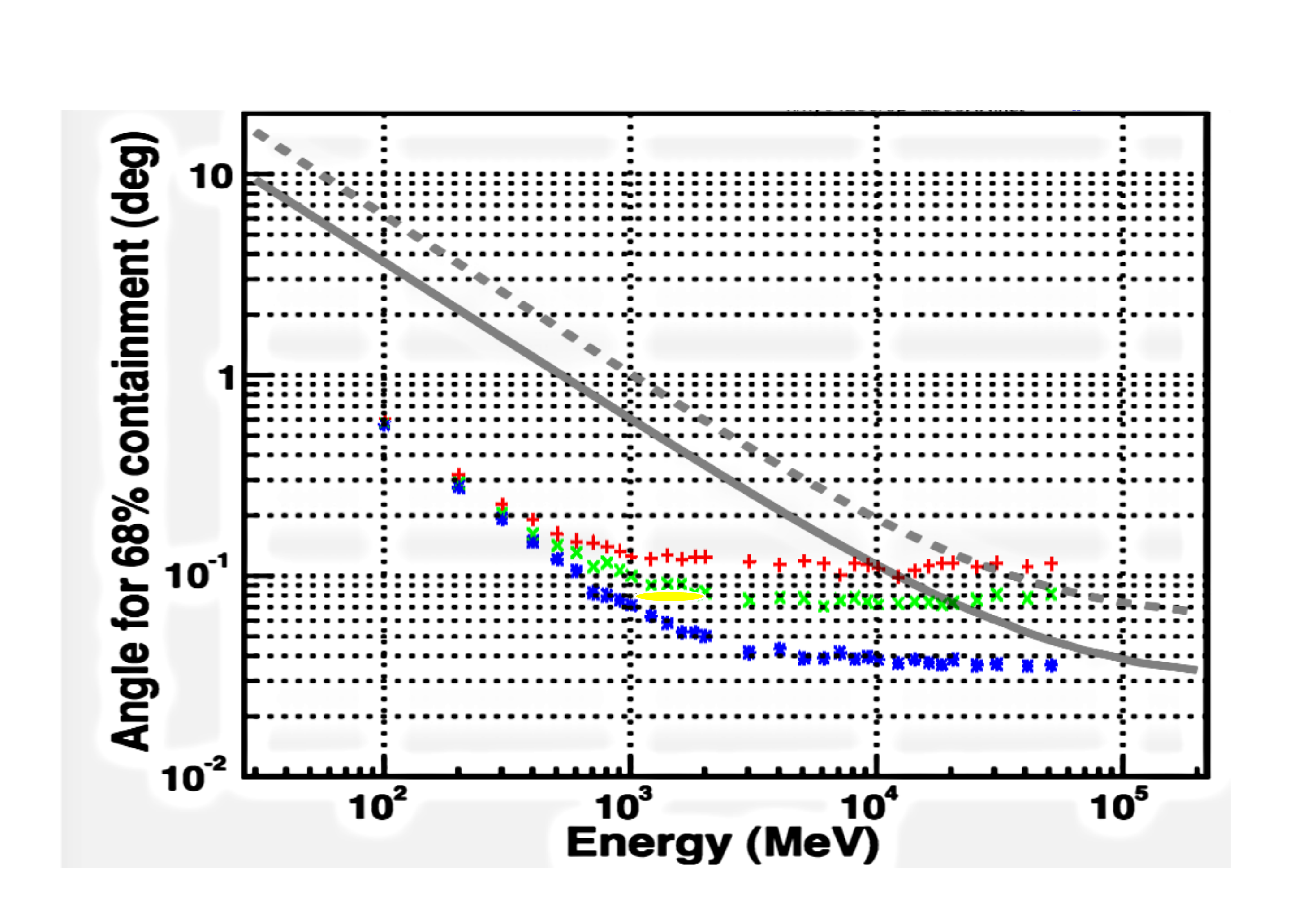}
\caption{The angular resolution for gamma-rays as a function of energy compared with Fermi LAT}
\label{AngularResolution}
\end{center}
\end{figure}

 Figure \ref{AngularResolution} shows the angular resolution for gamma-rays as a function of energy compared with Fermi LAT. Solid line and dashed line show the angular resolution of Fermi LAT. Plots with each symbol show the angular resolution of emulsion gamma-ray telescope for each readout accuracy (+:~0.3~$\mu$m, $\times$:~0.2~$\mu$m, $\ast$:~0.1~$\mu$m). A yellow point shows experimental data using accelerator gamma-ray beam. The emulsion gamma-ray telescope is much better than Fermi LAT and angular resolution was established experimentally.

\subsection{Energy range}
\begin{figure}[htbp]
\begin{center}
\includegraphics[scale=0.3]{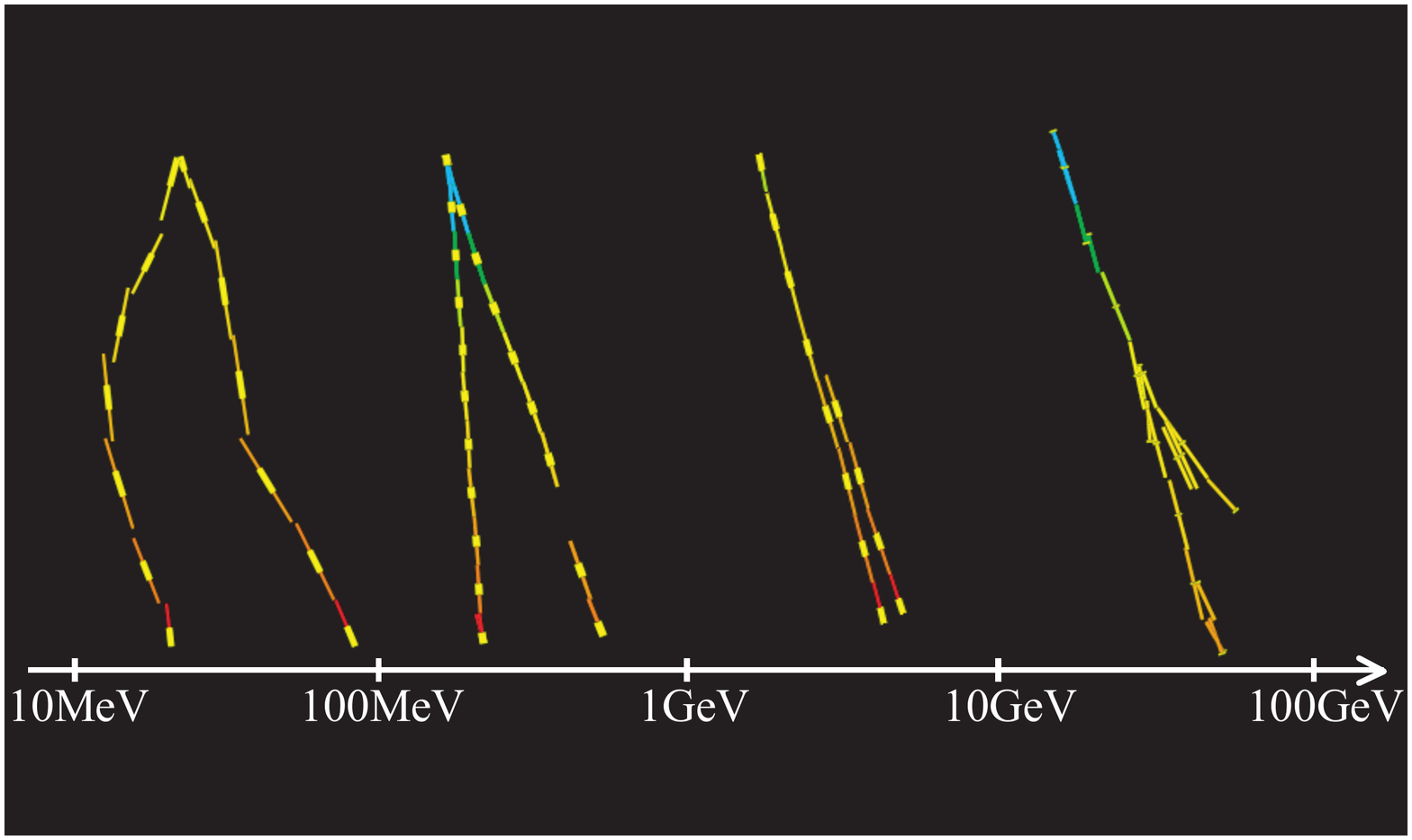}
\caption{Electron pairs for each gamma-ray energy detected with emulsion in test experiments}
\label{EnergyRange}
\end{center}
\end{figure}

 Systematic detection of electron pairs in emulsion is necessary for gamma-ray telescope. We did test experiments to establish systematic detection of electron pairs using accelerator gamma-ray beam (SPring-8\footnote{Inverse compton scattering gamma-ray beam, Maximum gamma-ray energy 2.4~GeV} and UVSOR\footnote{Inverse compton scattering gamma-ray beam, Maximum gamma-ray energy 47~MeV}) and atmospheric gamma-rays at mountain altitude (Mt. Norikura, 2770~m) \cite{BSproc}. Systematic detection of electron pairs was performed down to 50~MeV in these test experiments. Further study for systematic detection of electron pairs down to 10~MeV is going on. Figure \ref{EnergyRange} shows electron pairs for each energy detected in these test experiments.

\subsection{Polarization sensitivity}
 The modulation appears for the opening direction of electron pair created by linear polarized gamma-ray. By detecting the modulation, polarization is detected. Emulsion can detect the opening direction of electron pair by detecting beginning of electron pair. Thus emulsion is expected to have polarization sensitivity by collecting events. We are studying polarization sensitivity of emulsion by analyzing the sample exposed linear polarized gamma-ray by accelerator.

\subsection{Multi-stage shifter}
\label{MultiStageShifter}
 The attitude of balloon-borne telescope will change milliradian level per second. Thus milliradian angular resolution requires the time resolution below second. We developed multi-stage shifter as the time stamper \cite{MultiStageShifter}. The multi-stage shifter consists of emulsion films shifting at individual cycle. By combining track displacement for each stage, many independent states are created. It is more resolvable for longer time. Multi-stage shifter is similar to an analog clock which shows 12 hours with second accuracy by hands moving at different cycle. By increasing number of stages with a shorter cycle, the time resolution is improved. Multi-stage shifter achieves simple design, compact, light, high voltage free, low power and dead time free. We did the test experiment using cosmic rays on the ground to establish the time stamp method by multi-stage shifter. The time resolution of 1.5 second was obtained and the time stamp method by multi-stage shifter was established.

\section{The flight model}
\subsection{The first flight model}
\begin{figure}[htbp]
\begin{center}
\includegraphics[scale=0.25]{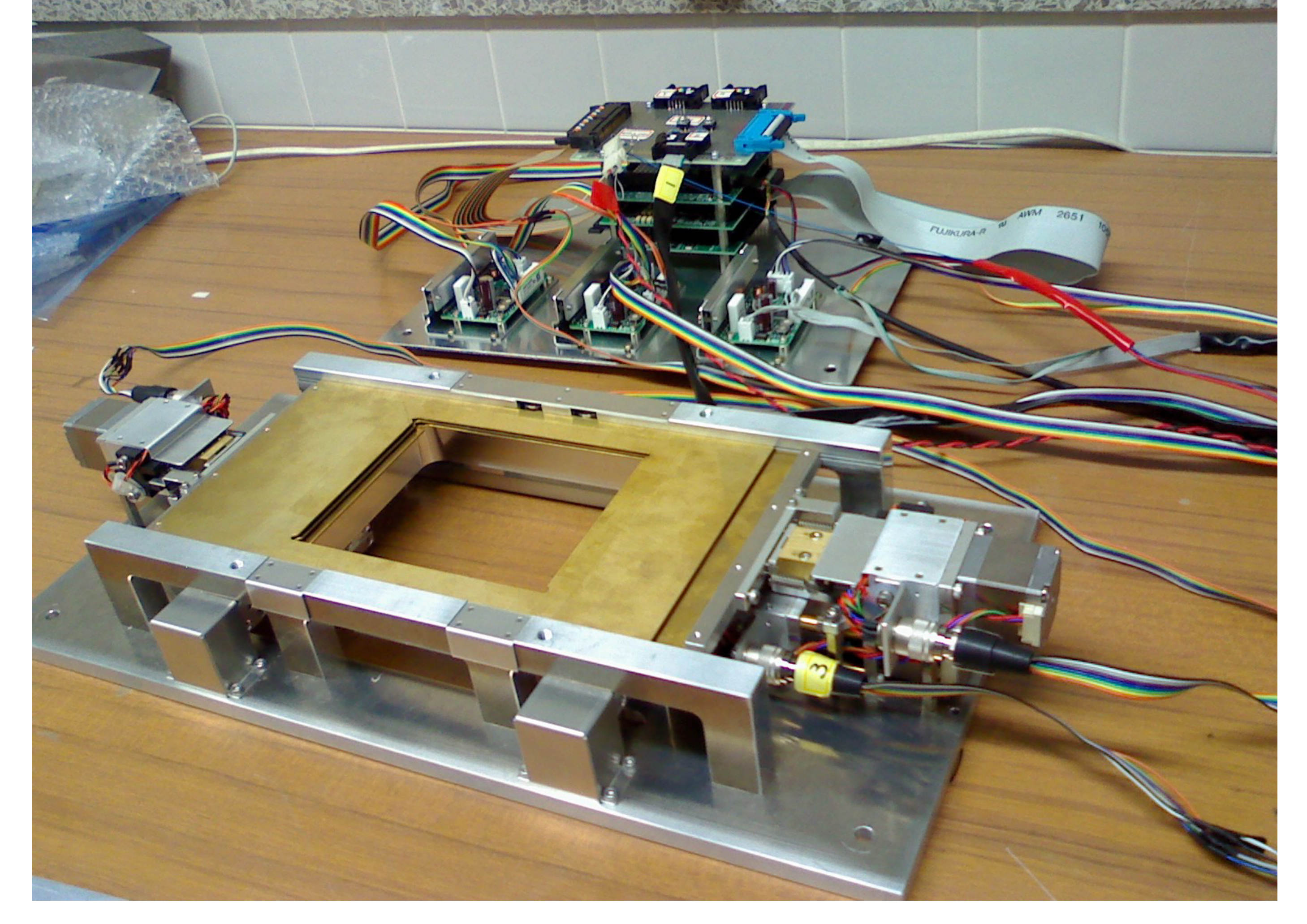}
\caption{The picture of the first flight model}
\label{1st}
\end{center}
\end{figure}

 Figure \ref{1st} shows the flight model of multi-stage shifter. We co-developed with Mitaka Kohki Co., Ltd. Aperture area is 12 cm $\times$ 10 cm. By setting emulsion films in the central space, with the converter above it and the calorimeter below, this becomes emulsion gamma-ray telescope (the first flight model).   Environmental test has been finished. The first flight model is ready for the flight.

\subsection{The second flight model}
\begin{figure}[htbp]
\begin{center}
\includegraphics[scale=0.275]{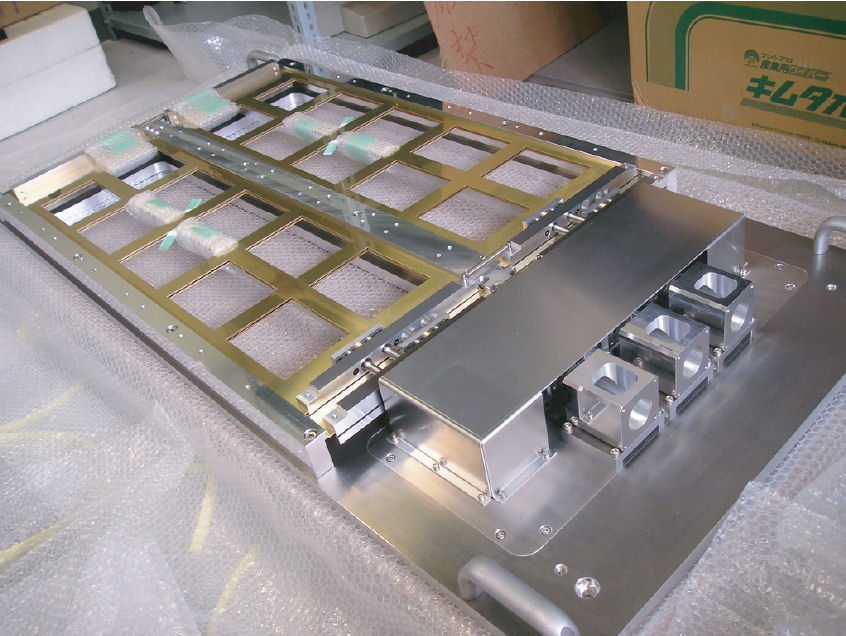}
\caption{The picture of the second flight model}
\label{2nd}
\end{center}
\end{figure}

 Figure \ref{2nd} shows the second flight model expanded from the first flight model. Aperture area is 12~cm $\times$ 10~cm $\times$ 20 units. The second flight model will be ready for the flight by next spring.

\section{Summary and outlook}
 By detecting beginning of electron pairs with emulsion, precise gamma-ray direction and gamma-ray polarization can be detected. With recent advancement in emulsion scanning system, emulsion analyzing capability is becoming powerful. By basic study, the perspective was obtained for the observation of cosmic gamma-ray with emulsion. We start the observation of cosmic gamma-ray by balloon flight with emulsion gamma-ray telescope. The first flight model is ready for  flight. With the first flight model, we test under the balloon flight environment with the flight duration above several hours and measure the background. The second flight model will be ready for the flight by next spring. With the second flight model, we observe known gamma-ray object with the flight duration above 6 hours and test with over all. With the future model, we start full scale observation by long duration flight and repeating flight.

\bigskip 
\begin{acknowledgments}
 We would like to thank the people who supported this work in various scenes. This work was supported by grants from Japan Society for the Promotion of Science, and grants for 21st Century COE Program and Global COE Program from the Ministry of Education, Culture, Sports, Science and Technology of Japan.
\end{acknowledgments}

\bigskip 

\end{document}